\documentclass[aip]{revtex4-1}
\usepackage{graphicx, amsmath, amssymb}

\begin{document}
 
\title{Sub-THz complex dielectric constants of montmorillionite clay thin samples with Na$^{+}$/Ca$^{++}$-ions}

\author{Rezwanur Rahman}
\affiliation{Department of Physics, Colorado School of Mines, Golden, CO 80401-1887, USA}
\affiliation{Department of Petroleum Engineering, Colorado School of Mines, Golden, CO 80401-1887, USA}
\affiliation{OCLASSH, Department of Petroleum Engineering, Colorado School of Mines, Golden, CO 80401-1887, USA}
\author{Douglas K. McCarty}
\affiliation{Chevron ETC, 3901 Briarpark, Houston, TX 77402, USA}
\author{Manika Prasad}
\affiliation{Department of Petroleum Engineering, Colorado School of Mines, Golden, CO 80401-1887, USA}
\affiliation{OCLASSH, Department of Petroleum Engineering, Colorado School of Mines, Golden, CO 80401-1887, USA}
\author{John A. Scales}
\affiliation{Department of Physics, Colorado School of Mines, Golden, CO 80401-1887, USA}

\begin{abstract}

    We implement a technique to
    characterize electromagnetic properties at frequencies 100 to 165
    GHz (3 cm$^{-1}$ to 4.95 cm$^{-1}$) of oriented montmorillionite
    samples  using an open cavity resonator connected
    to a sub-millimeter wave VNA (Vector Network Analyzer).
    We measured dielectric constants perpendicular to the
    bedding plane on oriented Na$^{+}$ and Ca$^{++}$-ion stabilized
    montmorillionite samples deposited on a glass slide at ambient
    laboratory conditions (room temperature and room light).  The clay
    layer is much thinner ($\sim$ 30 $\mu$m) than the glass
    substrate ($\sim$ 2.18 mm).  The real part of dielectric constant,
    $\epsilon_{re}$, is essentially constant over this frequency 
    range but is larger in
    Na$^{+}$- than in Ca$^{++}$-ioned clay.  The total electrical conductivity
    (associated with the imaginary part of dielectric constant,
    $\epsilon_{im}$) of both samples increases monotonically at lower
    frequencies ($<$ 110 GHz), but shows rapid increase
    for Na$^{+}$ ions in the regime $>$ 110 GHz. 
    The dispersion of the samples
    display a dependence on the ionic strength in the clay
    interlayers, i.e., $\zeta$-potential in the Stern layers.

\end{abstract}

\maketitle 

\section{Introduction}
Clay minerals have a complex layered structure with excahangeable
cations that can bind water molecules in the inter layers. With
increasing pressure and temperature, or in the presence of polar free
radicals, these interlayer cations can be exchanged. This cation
excahange capacity (CEC) of clay minerals affects their
fluid conductivity,\citep{RT86} and permeability\citep{RT86}; 
dielectric permittivity.\citep{WC82} Complex dielectric properties of clay are
crucial to determine hydrocarbon-contents in a oil-rich rocks. These
measurements are usually performed at logging frequencies in the 
kHz-range\citep{MM77, MC95} and between 0.5 MHz and 1.1
GHz.\citep{RT86} Complex conductivity of clayey materials between 1
mHz and 45 kHz for  CEC effects was characterized and
modeled.\citep{AR13} \citet{BC99} researched dielectric properties of
montmorillionite clay samples in detail,  explaining interlayer
polarization and relaxation mechanisms between 30 kHz-300 MHz. Some
clay minerals can swell due to hydration with water adsorbed in the
interlayer depends on the charge of the interlayer cations. Electrical
measurements can yield cation mobility. Dielectric measurements can be
instrumental to characterize the water absorbed in
montmorillionites\cite{WR68} The conductivities of montmorillionite
clay saturated by mono valent cations was studied where Debye and
Maxwell-Wagner relaxations are discussed in details.\citep{RC75, FJ65}
These frequency-and temperature-dependent measurements were in the
frequency range between 300 and 10,000 Hz, and from -150$^{\circ}$C to
+30$^{\circ}$C. THz dielectric constants of layered silicates
including muscovite, vermiculite, phlogopite, and biotite, have been
measured by THz-time domain (THz-TDS) spectroscopy.\citep{MJ09}
In this paper we look at Monmorillionite using much higher resolution CW methods 
based on harmonic multiplication of phase stabilized
microwaves, electronically generated; 
these methods provide very low noise/high dynamic range out to about 1.4 THz
at present.

At CSM We use 3 millimeter wave (or sub-THz) modalities: (1) A Quasi Optical
System,\citep{JS0689,JS0688,NG14} to study bulk properties, (2) A Near
Field Scanning System,\citep{MW09} to measure local properties, and
(3) the Open Hemispherical Cavity Resonator,\citep{RR13a,Dudorov95}
for samples that are too optically thin or low loss for quasi-optical
techniques.  Using cavity resonance
we measure these complex dielectric
constants of clay-thin films in 100-165 GHz or 0.1-0.16 THz and
investigate electrical properties in the presence of
Ca$^{++}$/Na$^{+}$-ions. We study how these cations influence
conductivity of free carriers, and relaxations. We also compare our
data with the low frequency measurements.\citep{RT86}
  
\section{Methods}
  We use an open hemispherical open cavity resonator with VNA (Vector
  Network Analyzer)\citep{RR13a} to measure electrical properties of
  thin sections of clay-samples with Ca$^{++}$/Na$^{+}$-ions
  infused. The cavity is a structure with two copper mirrors
  positioned at certain distance (the ``cavity length'') without any
  sidewalls. The top mirror is hemispherical and connected to two
  WR-10 waveguide couplers working as a transmitter and a receiver,
  and on the other hand the lower mirror is flat and smaller than the
  upper one in size. We measured the real part of refractive index of
  $\sim$ 1 mm-thick glass substrate (borosillicate) to be 1.98 at 310
  GHz which is the same as its theoretical value.\citep{RR13a} For
  details on the cavity and methodology see \citet{RR13a}.

\subsection{Open Cavity Resonator}
  The principle of this technique is cavity perturbation. The changes
  in axissymetric (00q) mode profiles, mainly the frequency-shift and
  linewidth-variation, between an empty cavity mode and the same mode
  in presence of a sample, determine the complex dielectric constant
  of the sample.\citep{cullen1,cullen2,TH96} The unloaded (empty)
  cavity has an axisymmetric mode spacing that is c/2L, where L is the distance
  between two mirrors, also known as cavity length. In our cavity,
  since L is around 15 cm, the unloaded mode spacing is about 1GHz.

  Putting a sample on the bottom mirror perturbs the modes in a
  calculable but nontrivial way. To avoid geometrical factors, we do a second
  perturbation which involves flipping the sample upside down. Since
  the boundary values of the E-field are different, we are able to get
  a simple (geometry-free) formula for the complex permittivity from 3
  sweeps around a 00q mode.

This complex dielectric permittivity is related to total electrical/optical 
conductivity and
absorption coefficient\citep{ST77} with the use of basic theory of
electromagnetism as
\begin{equation*}
n_{re} \alpha = 120 \pi \sigma_{re} = 30 \omega \epsilon_{im} 
\end{equation*}
Where n$_{re}$ and $\sigma_{re}$ are the real pert of the refractive
index and conductivity (in $\Omega^{-1}$cm$^{-1}$), $\epsilon_{im}$ is
imaginary part of the complex dielectric constant, and $\omega$ and
$\alpha$ are labeled for frequency and absorption coefficient, both of
them are in wave numbers (cm$^{-1}$), where 1cm$^{-1}$ = 30 GHz. The
conductivity is a macroscopic quantity which can be optical,
electronic or ionic depending on the system and the
frequency range of the probe. This is also valid for
absorption coefficient. These parameters, $\sigma_{re}$ and $\alpha$,
essentially describe the loss mechanisms in a material.

\subsection{Measurements}
   By sweeping the VNA, we identifie
d the axissymetric empty cavity
   modes based on constant frequency-spacing.  This fixes the cavity-length,
   l, to be 145.56 mm and kept it unchanged throughout the experiments. 
   We measured the eigenfrequency-shifts and modal quality factors 
   (Q-values related to a linewidth) for substrate-only, film up and
   film down positions in order to apply the differential method.  In
   order to determine uncertainty in the experiment, we repeat the
   entire procedure of inserting the sample, performing the
   measurements, and taking it out for six times and calculating the
   variations in frequency shifts and linewidth changes of
   substrate. In cavity paper,\citep{RR13a} we showed that by re-doing
   the entire procedure for six times for borosillicate glass
   substrate, we obtained standard deviations $<$ 1.0$\%$ in measuring
   its complex dielectric constant.  This uncertainty can also
   propagate to the calculations of dielectric constant.  We confirmed
   during each trial, the same part of the samples is probed to make
   it consistent.

\begin{figure}
  \includegraphics[width=70mm]{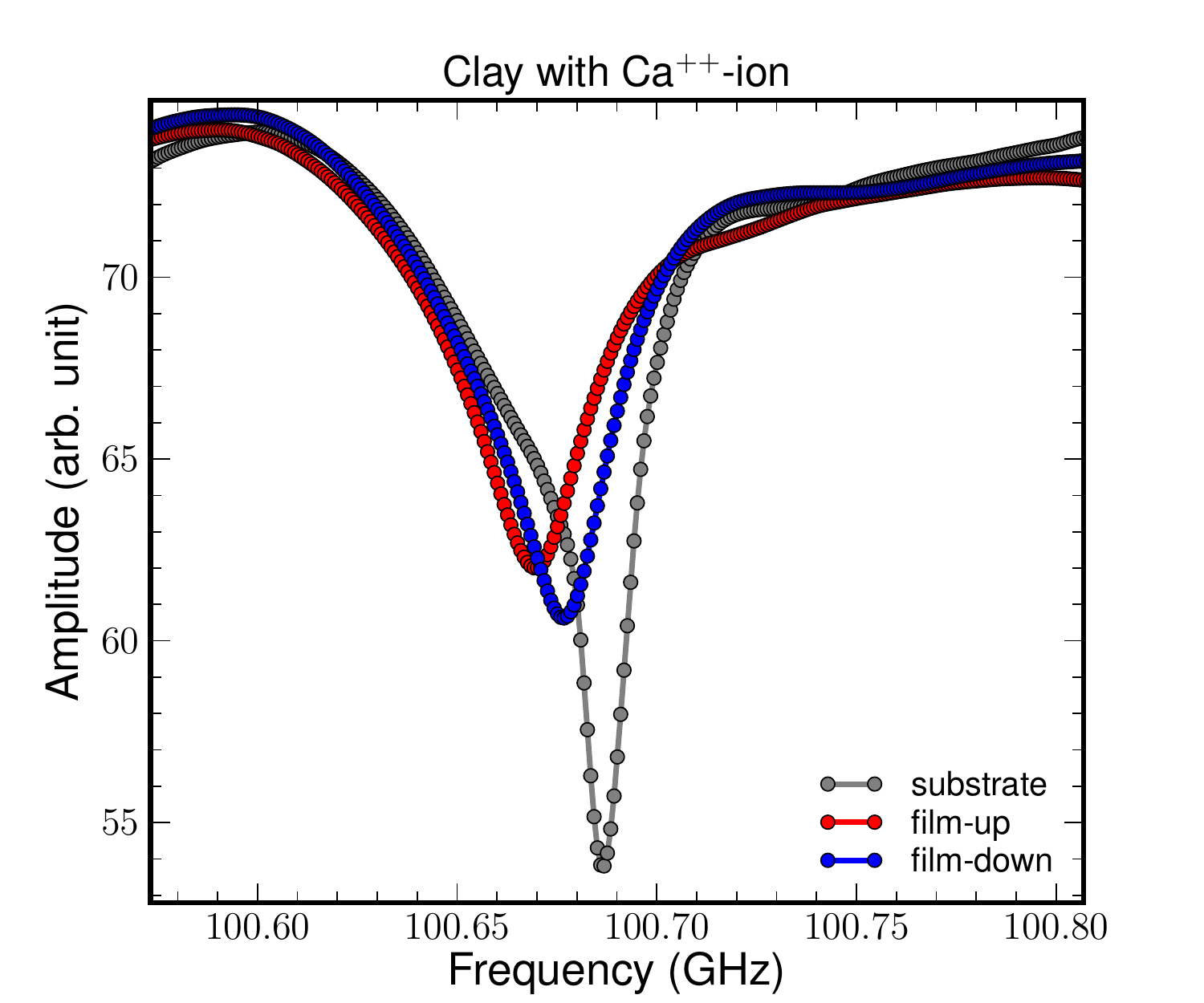}
  \label{fig:subf1}
  \includegraphics[width=70mm]{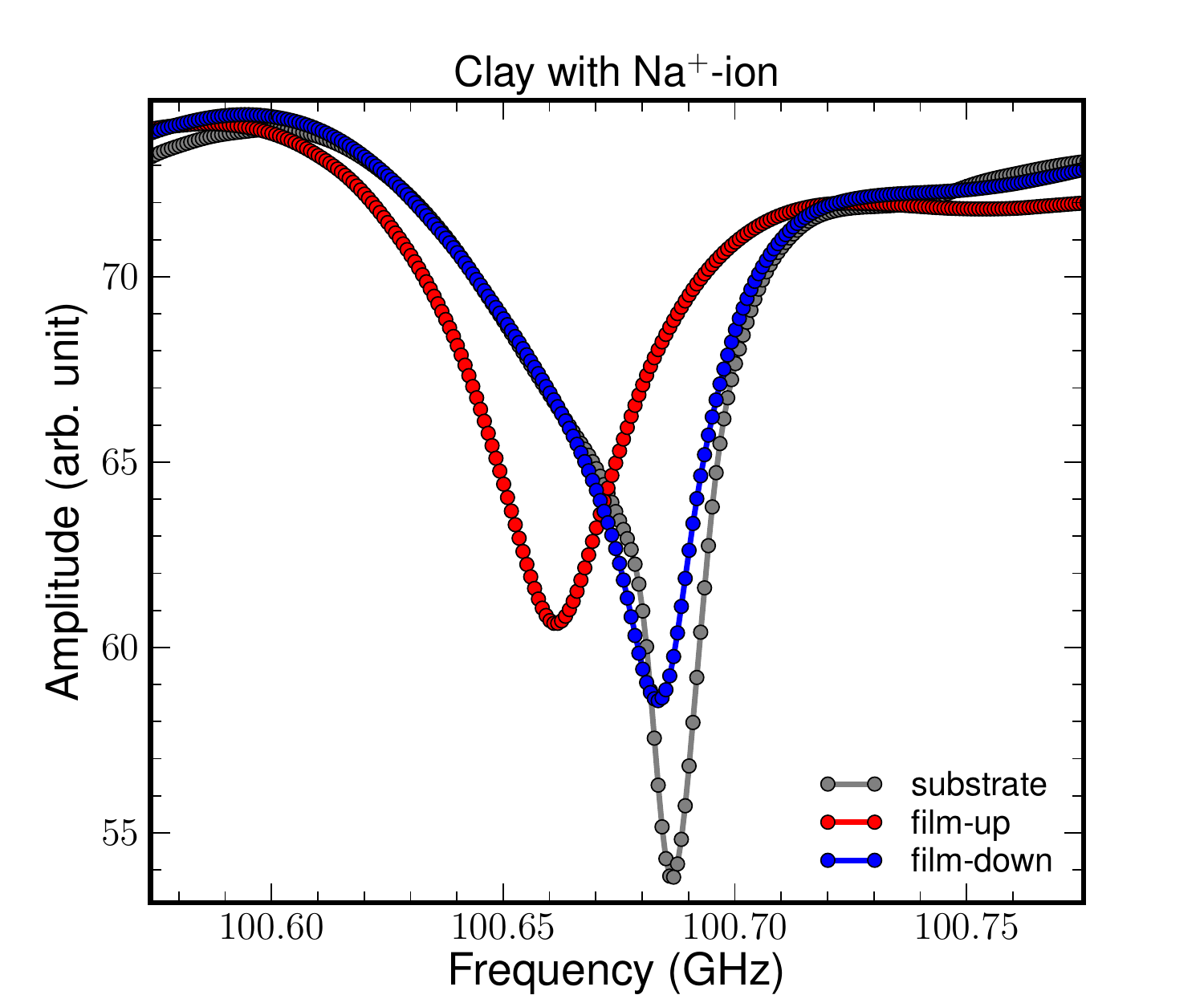}
   \label{fig:subf2}
   \caption{Perturbations with substrate only, film up, and film down set-ups
             for clay with (a) Ca$^{++}$ and (b)Na$^{+}$-ions (at room temperature}.
\label{fig:f12}
\end{figure}

  The frequency shifts due to film up, film down, and substrate only are used in the Eqs.(\ref{equ:flipfloptech}) for flipflop method, to
determine the real part of the refractive index of a thin film.\citep{RR13a,Dudorov95} 
\begin{equation}
\frac{\delta\nu_{f}}{\delta\nu_{s}} =  \frac{n_{f}^{2} - 1}{n_{s}^{2} - 1}.\label{equ:flipfloptech}
\end{equation}
With
\begin{equation}
 \delta\nu_{f} = \nu_{(fup)} - \nu_{(s)}.\label{fupshift}
\end{equation}
\begin{equation}
 \delta\nu_{s} = \nu_{(fdown)} - \nu_{(s)}.\label{fdownshift}
\end{equation}
Where, $\nu_{(fup)}$ , $\nu_{(fdown)}$ and $\nu_{(s)}$ represent the
eigenfrequency associated with the film on the top (film up), film at
the bottom (film down), and the substrate only, respectively.  The
term, $\delta\nu_{s}$, stand for the difference between the
eigenfrequencies associated with film at the bottom of the substrate
and the substrate only, and the term, $\delta\nu_{f}$, is the
difference between the eigenfrequencies with the film at the top of
substrate and the substrate only.  The required condition is the real
part of the refractive index of substrate must be known. The real of
part of the dielectric constant of thin film can obtained by
$\epsilon_{re}^{(f)}$=(n$_{re}^{(f)})^{2}$.  We need Q-values of film
up and substrate (only) to calculate the imaginary part of complex
dielectric constant of the thin film.  The Q-value of a resonant peak
(perturbation) is defined as Q=$\nu_{0}$/$\Delta\nu$. The Q-value is
related to the imaginary part of the refractive index by
n$_{im}$=1/2Q, so for a thin film, n$_{im}^{(f)}$=1/2Q$^{(f)}$. In the
fipflop method, Q$^{(f)}$=( Q$^{(film up)}$ - Q$^{(substrate)}$
). Now, we are able to compute the imaginary of the complex dielectric
constant of the thin film by $\epsilon_{im}^{(f)}$=2n$_{re}^{(f)}$
n$_{im}^{(f)}$.

\subsection{Samples and Sample preparations}
We studied montmorillionite clay minerals from Clay Mineral Society
that were treated to yield homoionic, univalent (Na$^{+}$) and
divalent (Ca$^{++}$) samples.  The samples were treated to remove
carbonate and iron oxide cements with a Na-acetate buffer and
Na-dithionite respectively.\citep{JML85} The sub-0.5 $\mu m$ equivalent
spherical diameter size fractions were separated from samples by
standard centrifugation methods.  Then the Na$^{+}$ saturated clay was
thoroughly cleaned with dialysis to remove excess salt.  To prepare
the Ca$^{++}$ exchanged smectite, a dialyzed Na$^{+}$ sample was
treated with a 1M solution of CaCl$_{2}$, and shaken for at least two
hours, excess solution was decanted and the process was repeated twice
more.  The excess Ca$^{++}$ salt was removed with
dialysis.\citep{MDM97} Oriented aggregates were made by evaporation
onto glass slides to provide a sample $\sim$ 4 cm long with at least
10 mg clay per cm$^{2}$.\citep{MDM97}

In addition to millimeter wave EM analysis,
we performed a variety of measurements on the samples, including
thermal gravimetric analysis (TGA), subcritical Nitrogen gas adsorption
(SGA) and X-ray diffraction (XRD) on oriented samples

TGA experiments were performed using a {\it TA
Instruments Q5000} purged with dry filtered reagent grade nitrogen.
Prior to the analysis, the samples were equilibrated at ambient
conditions (~50\% RH).  Approximately 25 mg of each sample was used
for the experiments.  The weighing error is believed to be $<$0.001 mg.
The heating rate for all cycles was fixed at 5 $^\circ$C/min.  The inert
nitrogen purge gas flow rate was constant at 25 mL/min.  DTG patterns
reflecting reaction rates and relative hydration energy were obtained
by taking the first derivative of the percent weight loss
vs. temperature curves.

Specific surface area (SSA), and pore-size distribution (PSD), of the
Na$^+$ and Ca2$^+$ smectite forms were measured using the subcritical
nitrogen gas adsorption (SGA) at 77 K.  About 1-2 g of sample was
degassed by heating at 200 C under vacuum (10 mmHg) until the
out-gassing rates was $<$2 mmHg/min over a 15 min interval.
Measurements in both adsorption and desorption mode were performed
over the entire partial pressure range with average number of 85
measurement points.  SSA for each sample was determined by inversion
of the adsorption branch of the isotherm using a modified BET
procedure \citep{Rouquerol2007}.  Pore-size distribution (PSD) is obtained
by inverting the adsorption branch of the isotherm using
Barett–Joyner–Halenda (BJH) method assuming cylindrical nonconnecting
pores \cite{Kuila2014}.

The sub 500 nm  equivalent spherical diameter size fraction was separated
from bulk material by standard centrifugation methods following
treatment to remove carbonate and Fe-oxide cements with a Na-acetate
buffer and Na-dithionite, respectively (Jackson 1985).  A portion of
the Na+ saturated clay was treated by dialysis to remove excess salt
and retained, and a second portion was thoroughly exchanged with Ca2+,
and also treated with dialysis to remove excess salt (McCarty et al.,
2009).

Oriented aggregates were made by evaporation onto glass slides to
provide a sample ~4 cm long with at least 15 mg clay per cm2 (Moore
and Reynolds 1997).  Diffraction scans were collected with a Thermo
Xtra diffractometer with a θ-θ goniometer and a 250 mm radius,
equipped with a solid-state Si detector in the air-dried (AD) state,
and after ethylene glycol (EG) treatment by vapor solvation in a
heated chamber (60 °C).  The scans were made from 2 to 52 °2θ with a
0.02 °2θ step increment and counting rate of 4 s per step or longer
using CuKα radiation transmitted through a 1.00 mm divergence and 1.80
mm antiscatter slit.  Detector slits were 2.00 and 0.3 mm.  The
dielectric analysis was performed on these oriented aggregate
preparations.

Finally XRD patterns from the air-dried oriented aggregate smectite
specimens in Ca2+ and Na+ forms were simulated to reveal the relative
proportions of interlayer spacings corresponding to 2, 1, and 0 water
layers of interlayer cation hydration\citep{drits1976,drits1997,Sahkarov1999,McCarty2009}.

\section{Results and Discussion}
  The real part of the dielectric constants of both clay samples
  maintain (almost) constant values indicating that carrier
  concentrations are low in both.\citep{ZLN83,RR14} The imaginary part
  of dielectric constant and electrical conductivity of Na$^{+}$-ionized one,
  increase at two different rate where the faster in the higher
  frequency range and slower at low frequencies. The faster rate
  represents a nonlinear increase.  The presence of both Debye
  relaxation\citep{PD29}, and Maxwell-Wagner relaxation\cite{BH53} at
  radio frequencies was reported.\citep{RC75}
  But between 100 and 165 GHz, phonon
  induced-relaxation is dominant.\citep{ZLN83}

\begin{table*}
\caption{\label{table1}Dielectric parameters for clay smaples with Na$^{+}$ and Ca$^{++}$-ions from various sources:\\
                       ($^{*}$ indicates the data presented in this research paper)\\
                       ($^{**}$ mentions these values are normalized with water)}
\begin{ruledtabular}
\begin{tabular}{cccccc}
 \multicolumn{2}{c}{Clay-Na$^{+}$}  &\multicolumn{2}{c}{Clay-Ca$^{++}$}                       &Frequency                    &Sources\\
         \cline{1-2}                        \cline{3-4}                                                                   
 $\epsilon_{re}$&$\sigma_{re}^{*}$($\Omega^{-1}$ m$^{-1}$)& $\epsilon_{re}$&$\sigma_{re}^{*}$($\Omega^{-1}$ m$^{-1}$)& (kHz/MHz/GHz)                                        &(Data)        \\ \hline
 $\sim$ 217 - 80  &$\sim$ 0.009 - 0.055           &$\sim$ 104 - 80         &$\sim$ 0.0046 - 0.02    &0.5 MHz -0.1 GHz        &Raythatha et al.\citep{RT86} \\
  7.8             &0.69 - 40.2                     &4.86         &0.48 - 4.98                       &100 - 165 GHz           &Rahman et al.$^{*}$\\
       5 - 6$^{**}$      &                    &          & 0.0015 - 0.00375$^{**}$                   &1 GHz                            &Well logging\citep{NS11}\\
\end{tabular}
\end{ruledtabular}
\end{table*}

\begin{figure*}
  \includegraphics[width=60mm]{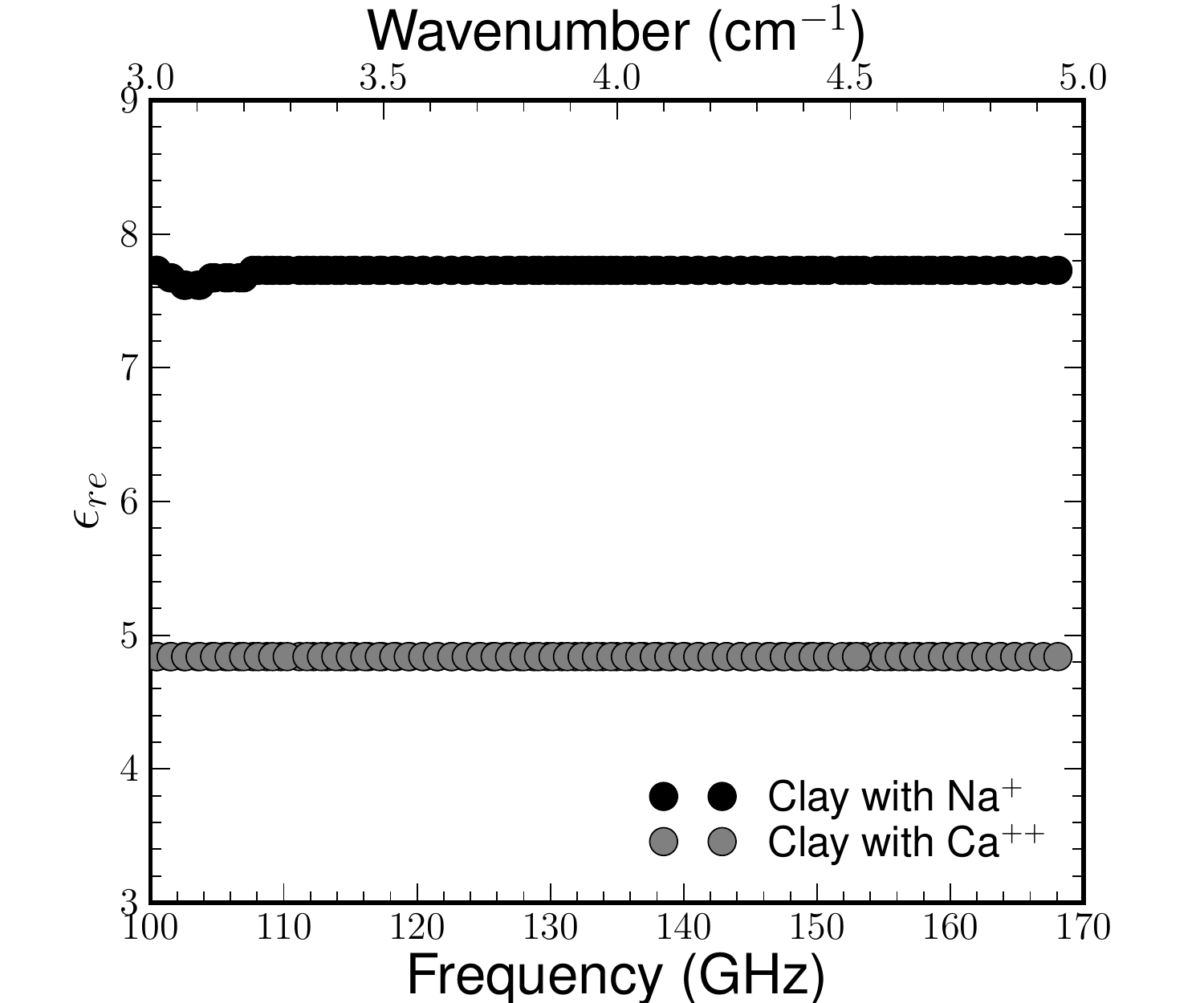}
  \label{fig:subf3}
\includegraphics[width=60mm]{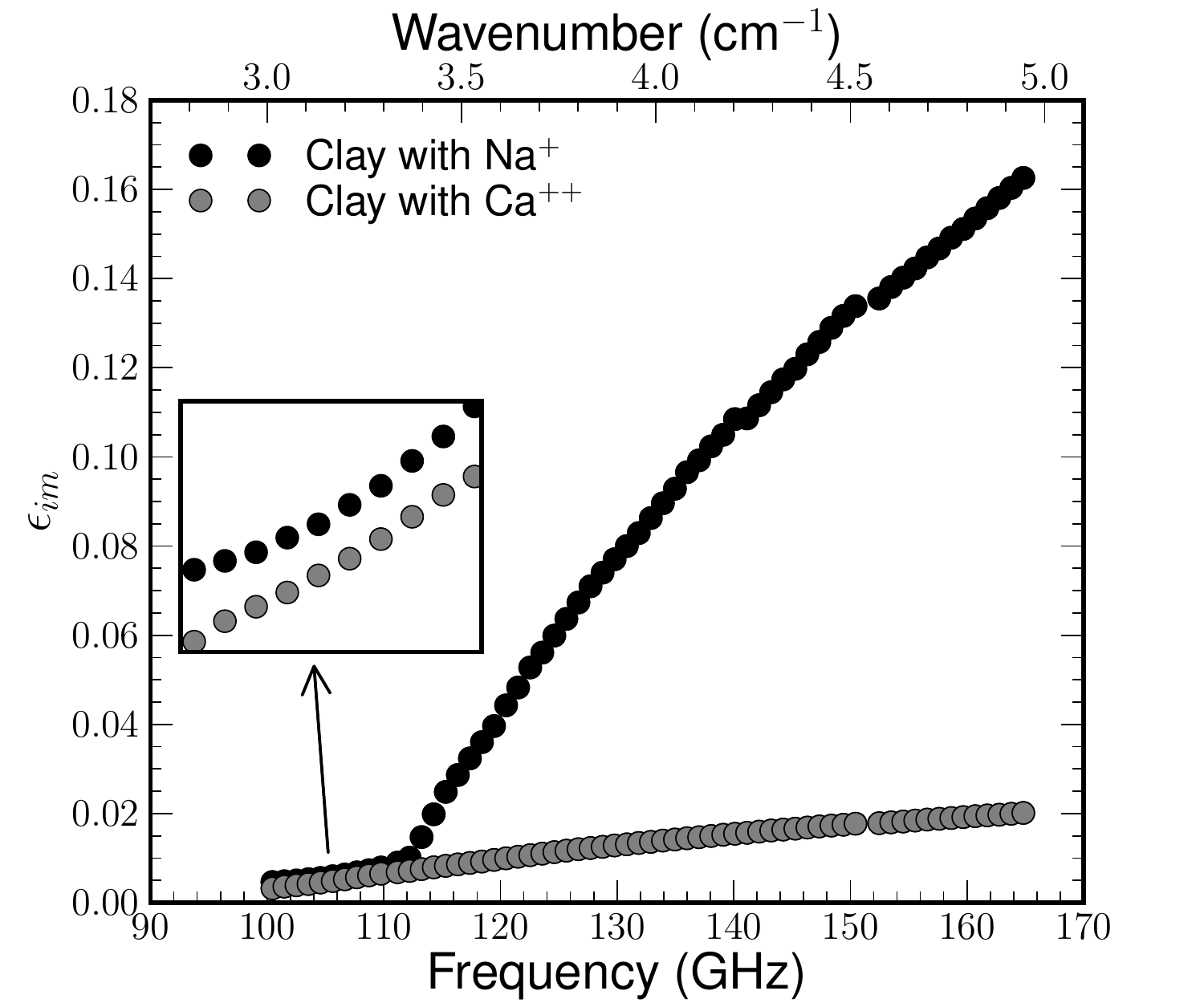}
   \label{fig:subf4}
\includegraphics[width=60mm]{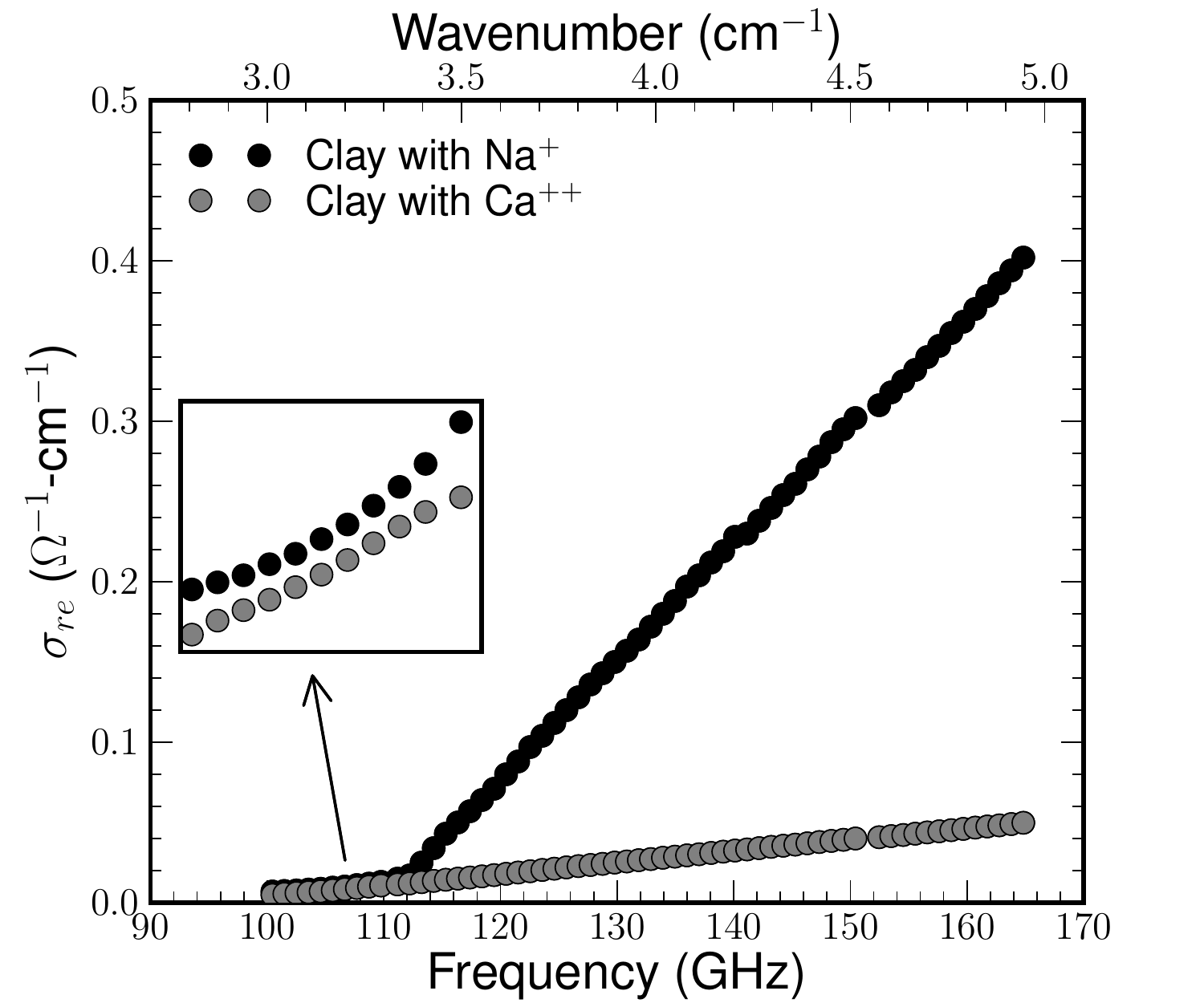}
   \label{fig:subf5}
 \caption{High frequency (100-165 GHz) data of (a) real part and (b)
   imaginary part of the complex dielectric constants of clay samples
   with Na$^{+}$/Ca$^{++}$-ions (at room temperature); (c)
   conductivity of of both samples (inner plot is the expansion of the
   100-110GHz responses).}
\label{fig:f345}
\end{figure*}

\begin{figure}
 \centerline{
     \includegraphics[width=140mm]{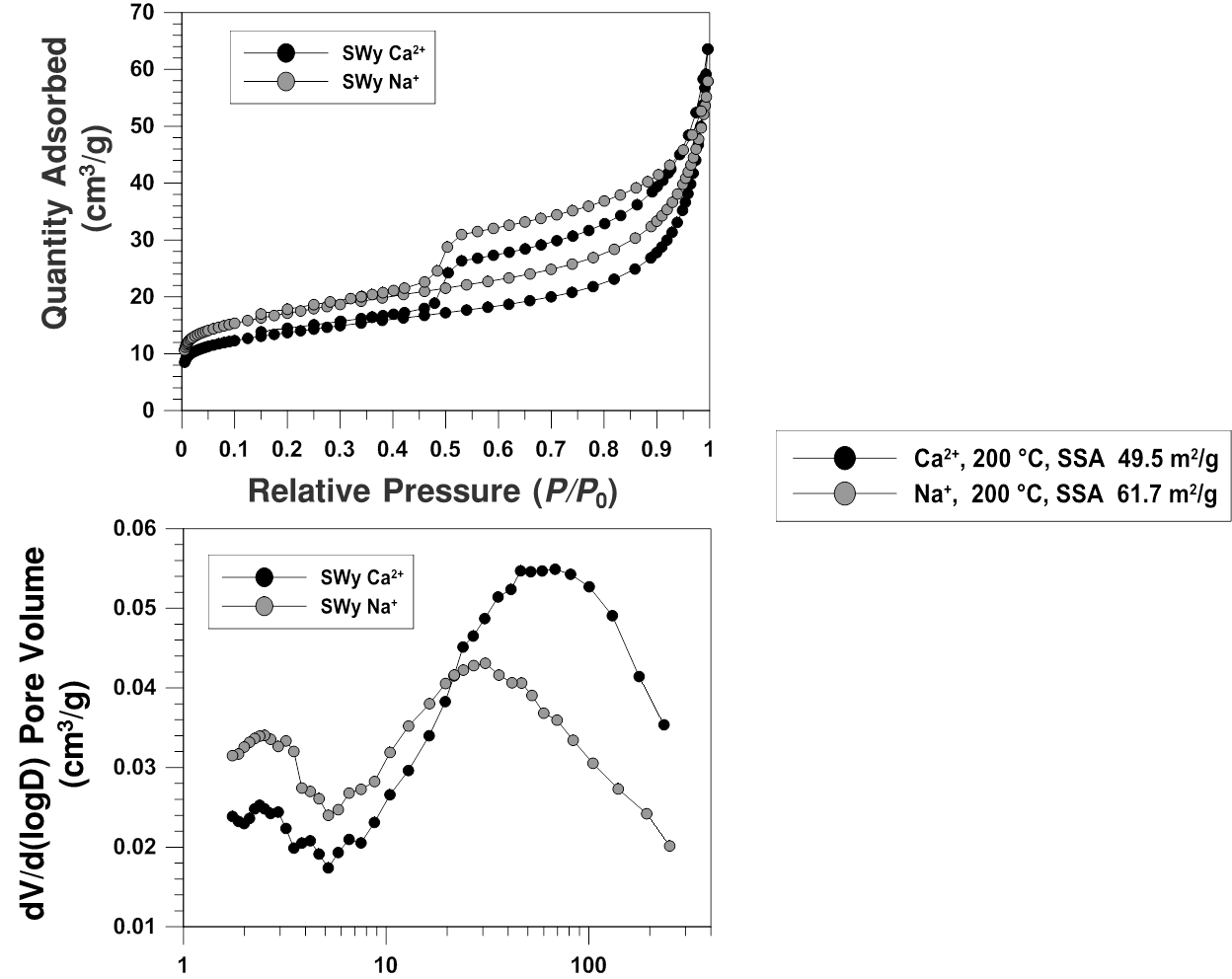}}
     \caption{Thermal data for Na$^{+}$ (shaded line)  and Ca$^{++}$ (dark
line). forms of SWy smectite used in this study.  (A) Weight loss (\%) vs. temperature, and (B), DTG reaction rate curves vs. temperature (see text for details).}
 \label{Isothermdata_NaCa}
\end{figure} 

\begin{figure}
 \centerline{
     \includegraphics[width=100mm]{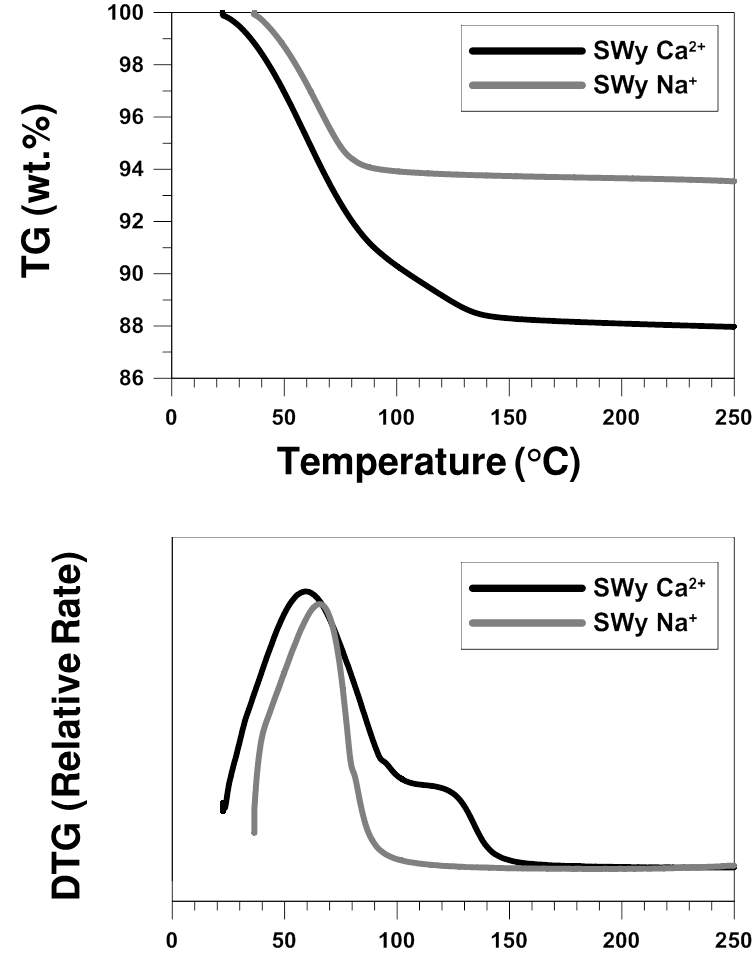}}
   \caption{Subcritical nitrogen gas adsorption (SGA) data at 77 K for
     the Ca$^{2+}$ (black points) and Na$^+$ (shaded points) forms of SWy
     smectite.  (A) Adsorption and desorption isotherms with relative
     pressure (P/P0) compared with quantity of adsorbed gas. (B) Pore
     size distribution of each sample, where pore diameter is compared
     to the volume of each size.  Specific surface area, and degas
     temperature of each sample is shown in the legend (see text).}
 \label{TGA-DTG_NaCa}
\end{figure}

 \citet{AR13} proposed a theoretical model to study complex
  conductivity-dependences on cation excahange capacity(CEC), specific
  surface area (SSA), and salinity for clay samples at low frequency.
  This model also relates CEC to SSA area with
  consistency.\citep{AR13} It is reported that SSA increases
  exponentially for Na$^{+}$-clay and linearly for Ca$^{++}$-clay
  sample.\citep{KE81} From our research, it is evident that
  conductivity depends on CEC or SSA. The larger CEC or
  expnonetial-growing SSA can contribute to more disperse Stern layer
  and the smaller CEC or linearly-progressive SSA stabilize the Stern
  layer.  Thus, electrical conductivity is linked to CEC (Cation Exchange
  Capacity) and zeta potential in the Stern layer.\citep{MC95} The
  Na$^{+}$ makes a thicker unstable double layer where these high
  mobility ion are able to polarize rapidly. Therefore, the
  conductivity is more dispersive. This interlayer
  polarization is correlated to relaxation mechanisms.  The relaxation
  processes is, therefore, dependent of $\zeta$-potential which is
  also correlated to CEC.\citep{DZ10} On the other hand, The imaginary
  part of dielectric constant and conductivity of the sample with
  Ca$^{++}$ increase monotonically and sublinearly.  The Ca$^{++}$
  creates more stable double Stern layer. Due to low mobility, the
  interlayer polarization is less disperse so its conductivity is
  sublinear. Since, $\epsilon_{im}$ = 2 (1/$\nu$) $\sigma_{re}$, the
  effect of (1/$\nu$) is more into imaginary part of complex
  dielectric constant of samples with Na$^{++}$ than that of
  Ca$^{++}$. In this case, $\epsilon_{im}$ at high frequencies faced
  steeper decrease than at lower frequency ends.

From Table (\ref{table1}), it is obvious that in the sub-THz, the
conductivities of clay-Na$^{+}$/Ca$^{++}$-ions are almost 2 to 3
orders of magnitude higher than in the RF/microwave range; on the
contary, the real part of dielectric constants for both samples,
decrease by $\sim$ one order of magnitude.  This illustrates clay
sample with Na$^{+}$ may have multiple relaxations. These high
$\sigma_{re}$ values (for both ions) indicate the electron
polarization in the Stern layer involve with phonon-mediated
interactions. The lower and constant values of $\epsilon_{re}$ also
confirm that there are depletations of mobile charges due to phonon
interactions causing the diffused layers to thin.

\begin{figure}
 \centerline{
     \includegraphics[width=100mm]{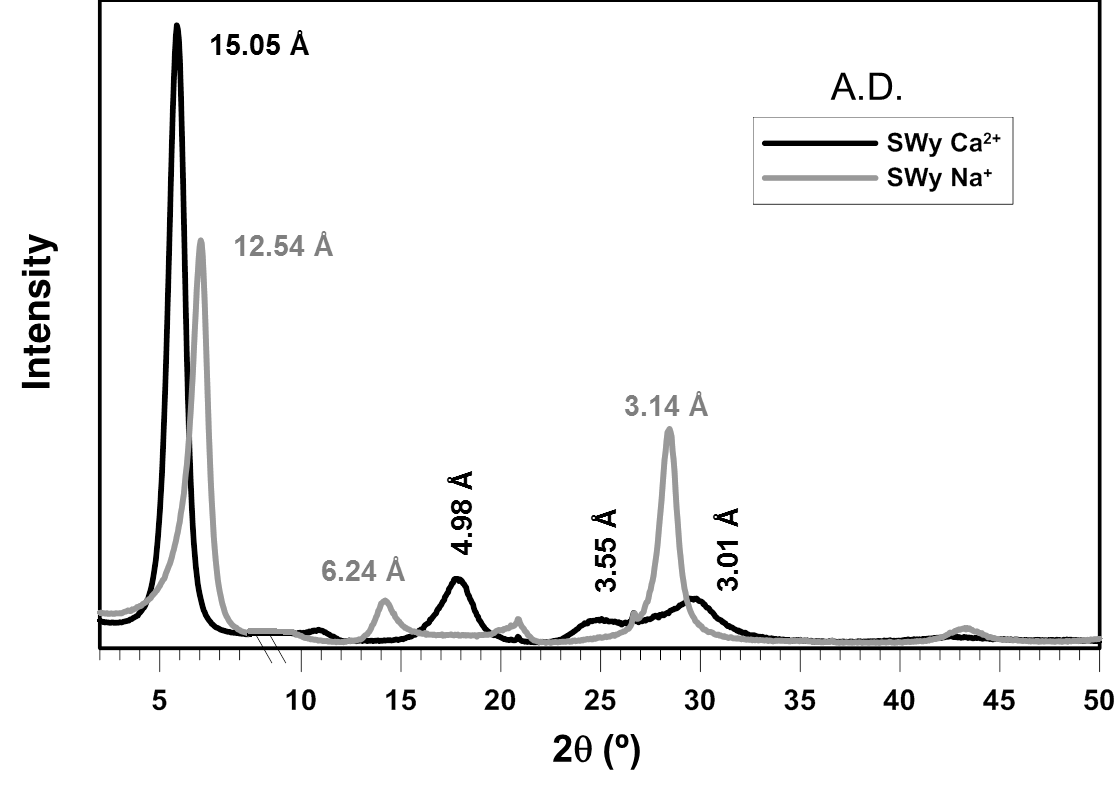}}
   \caption{X-ray diffraction patterns collected in the air-dried (AD) ambient state from oriented aggregate sample preparations comparing Ca2+ (black) and Na+ (shaded) smectite forms.  Basal 00l reflection d-spacings are shown on the figure in angstrom units (1\AA = 0.1 nm).}
 \label{XRD1}
\end{figure}

\begin{figure}
 \centerline{
     \includegraphics[width=100mm]{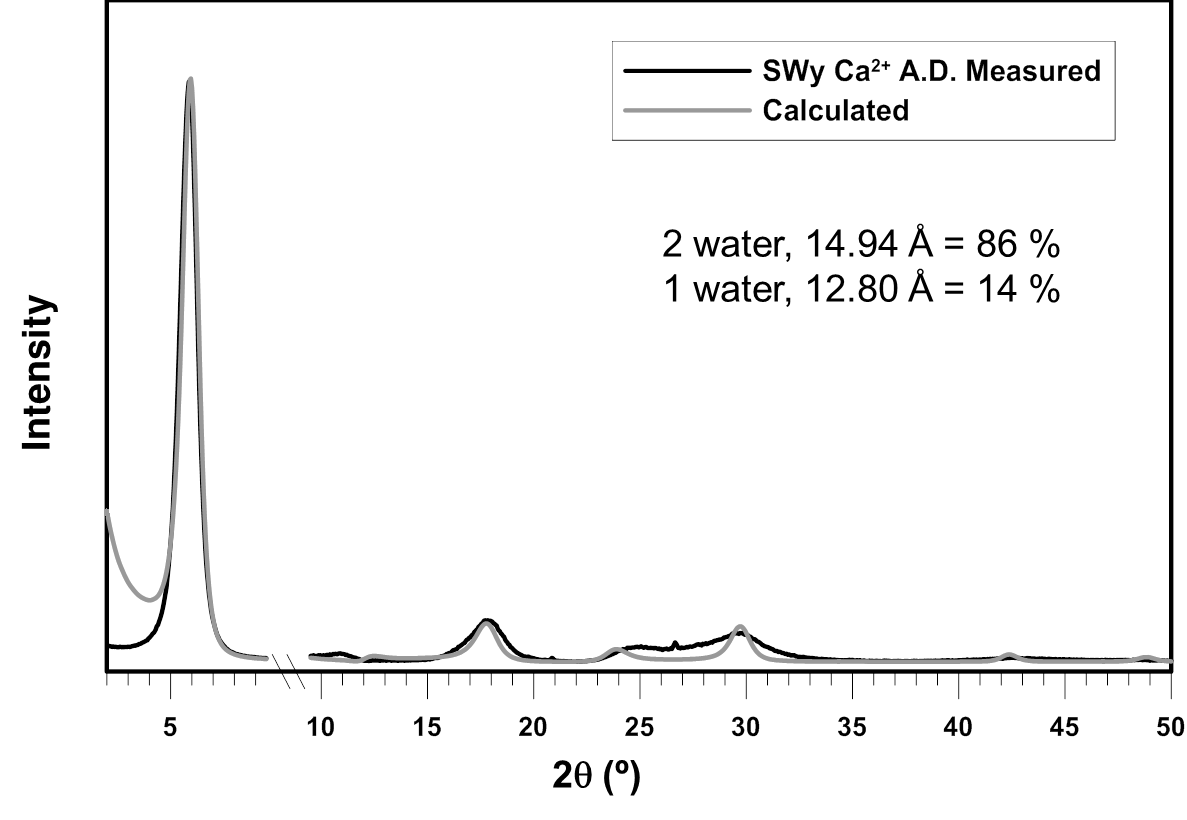}}
   \caption{ Experimental (black) X-ray diffraction pattern collected in the air-dried (AD) state from the Ca2+ form oriented aggregate smectite sample preparation compared with the simulated diffraction pattern (shaded).  The simulated diffraction pattern consists of 86\% of smectite layers with 2 layers of water molecules having a d-spacing of 14.94 \AA, and 14\% of smectite layers having 1 layer of water molecules and a d-spacing of 12.80 \AA (1\AA = 0.1 nm).}
 \label{XRD2}
\end{figure}

\begin{figure}
 \centerline{
     \includegraphics[width=100mm]{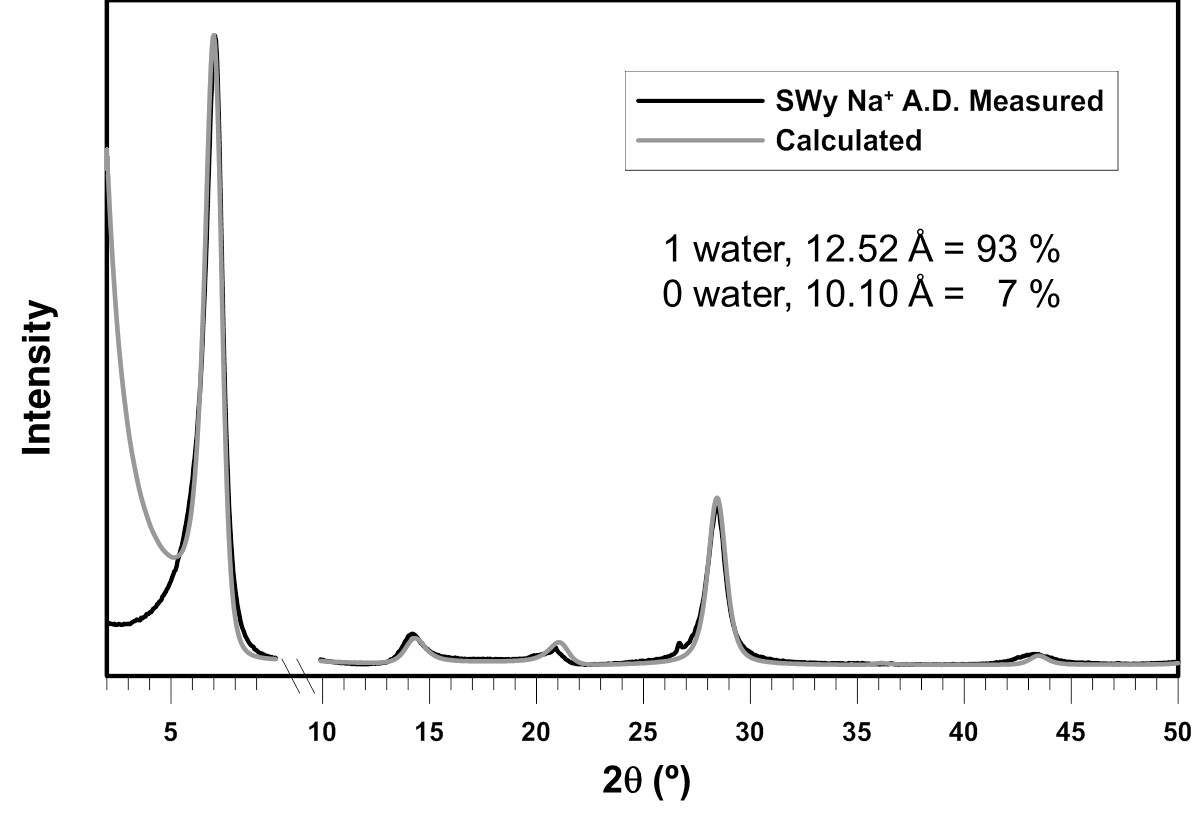}}
   \caption{Experimental (black) X-ray diffraction pattern collected in the air-dried (AD) state from the Na$^+$ form oriented aggregate smectite sample preparation compared with the simulated diffraction pattern (shaded).  The simulated diffraction pattern consists of 93\% of smectite layers with 1 layer of water molecules having a d-spacing of 12.52 \AA, and 7\% of smectite layers having 0 layers of water molecules and a d-spacing of 10.10 \AA (1\AA = 0.1 nm).}
 \label{XRD4}
\end{figure}

Finally we looked at the measured and simulated XRD patterns collected for the various samples (Figures \ref{XRD1}, \ref{XRD2} and \ref{XRD4}).

\clearpage

\section{Conclusion}
  We measured sub-THz complex dielectric properties of clay samples
  with Na$^{+}$/Ca$^{++}$, and compute their $\epsilon_{re}$ and
  $\sigma_{re}$.  We illustrate the connections between
  electromagnetic parameters and components of surface chemistry such
  as CEC or SSA, $\zeta$-potential.  This also enables us to study the
  clay content and free radicals in shales, and to investigate their
  CEC and $\zeta$-potential-dependences.  In future, we will study
  more different ionized clay samples and broader frequency ranges to
  capture more complete frequency-dependences of $\epsilon_{re}$ and
  $\epsilon_{im}$. This will allow us to model computationally the
  dispersions of these parameters.

\section{Acknowledgement}

This work was supported by OCLASSH consortium and the US Department of Energy
(Basic Energy Science) under grant DE-FG02-09ER16018.   This work is dedicated to the memory of Mike Batzle, who inspired us all.

\bibliographystyle{apsrev4-1}
\bibliography{main}

\begin{thebibliography}{35}%
\makeatletter
\providecommand \@ifxundefined [1]{%
 \@ifx{#1\undefined}
}%
\providecommand \@ifnum [1]{%
 \ifnum #1\expandafter \@firstoftwo
 \else \expandafter \@secondoftwo
 \fi
}%
\providecommand \@ifx [1]{%
 \ifx #1\expandafter \@firstoftwo
 \else \expandafter \@secondoftwo
 \fi
}%
\providecommand \natexlab [1]{#1}%
\providecommand \enquote  [1]{``#1''}%
\providecommand \bibnamefont  [1]{#1}%
\providecommand \bibfnamefont [1]{#1}%
\providecommand \citenamefont [1]{#1}%
\providecommand \href@noop [0]{\@secondoftwo}%
\providecommand \href [0]{\begingroup \@sanitize@url \@href}%
\providecommand \@href[1]{\@@startlink{#1}\@@href}%
\providecommand \@@href[1]{\endgroup#1\@@endlink}%
\providecommand \@sanitize@url [0]{\catcode `\\12\catcode `\$12\catcode
  `\&12\catcode `\#12\catcode `\^12\catcode `\_12\catcode `\%12\relax}%
\providecommand \@@startlink[1]{}%
\providecommand \@@endlink[0]{}%
\providecommand \url  [0]{\begingroup\@sanitize@url \@url }%
\providecommand \@url [1]{\endgroup\@href {#1}{\urlprefix }}%
\providecommand \urlprefix  [0]{URL }%
\providecommand \Eprint [0]{\href }%
\providecommand \doibase [0]{http://dx.doi.org/}%
\providecommand \selectlanguage [0]{\@gobble}%
\providecommand \bibinfo  [0]{\@secondoftwo}%
\providecommand \bibfield  [0]{\@secondoftwo}%
\providecommand \translation [1]{[#1]}%
\providecommand \BibitemOpen [0]{}%
\providecommand \bibitemStop [0]{}%
\providecommand \bibitemNoStop [0]{.\EOS\space}%
\providecommand \EOS [0]{\spacefactor3000\relax}%
\providecommand \BibitemShut  [1]{\csname bibitem#1\endcsname}%
\let\auto@bib@innerbib\@empty
\bibitem [{\citenamefont {Raythatha}\ and\ \citenamefont {Sen}(1986)}]{RT86}%
  \BibitemOpen
  \bibfield  {author} {\bibinfo {author} {\bibfnamefont {R.}~\bibnamefont
  {Raythatha}}\ and\ \bibinfo {author} {\bibfnamefont {P.~N.}\ \bibnamefont
  {Sen}},\ }\href@noop {} {\bibfield  {journal} {\bibinfo  {journal} {Journal
  of Colloid and Interface Science}\ }\textbf {\bibinfo {volume} {109}}
  (\bibinfo {year} {1986})}\BibitemShut {NoStop}%
\bibitem [{\citenamefont {Chew}\ and\ \citenamefont {Sen}(1982)}]{WC82}%
  \BibitemOpen
  \bibfield  {author} {\bibinfo {author} {\bibfnamefont {W.~C.}\ \bibnamefont
  {Chew}}\ and\ \bibinfo {author} {\bibfnamefont {P.~N.}\ \bibnamefont {Sen}},\
  }\href@noop {} {\bibfield  {journal} {\bibinfo  {journal} {Journal of
  Chemical Physics}\ }\textbf {\bibinfo {volume} {77}},\ \bibinfo {pages} {2042
  } (\bibinfo {year} {1982})}\BibitemShut {NoStop}%
\bibitem [{\citenamefont {Mehran}\ and\ \citenamefont
  {Arulanandan}(1977)}]{MM77}%
  \BibitemOpen
  \bibfield  {author} {\bibinfo {author} {\bibfnamefont {M.}~\bibnamefont
  {Mehran}}\ and\ \bibinfo {author} {\bibfnamefont {K.}~\bibnamefont
  {Arulanandan}},\ }\href@noop {} {\bibfield  {journal} {\bibinfo  {journal}
  {Clay and Clay Minerals}\ }\textbf {\bibinfo {volume} {25}},\ \bibinfo
  {pages} {39 } (\bibinfo {year} {1977})}\BibitemShut {NoStop}%
\bibitem [{\citenamefont {Chorom}\ and\ \citenamefont
  {Rengasamy}(1995)}]{MC95}%
  \BibitemOpen
  \bibfield  {author} {\bibinfo {author} {\bibfnamefont {M.}~\bibnamefont
  {Chorom}}\ and\ \bibinfo {author} {\bibfnamefont {P.}~\bibnamefont
  {Rengasamy}},\ }\href@noop {} {\bibfield  {journal} {\bibinfo  {journal}
  {European Journal of Soil Science}\ }\textbf {\bibinfo {volume} {46}},\
  \bibinfo {pages} {657 } (\bibinfo {year} {1995})}\BibitemShut {NoStop}%
\bibitem [{\citenamefont {Revil}\ \emph {et~al.}(2013)\citenamefont {Revil},
  \citenamefont {Eppehimer}, \citenamefont {Skold}, \citenamefont {Karaoulis},
  \citenamefont {Godinez},\ and\ \citenamefont {Prasad}}]{AR13}%
  \BibitemOpen
  \bibfield  {author} {\bibinfo {author} {\bibfnamefont {A.}~\bibnamefont
  {Revil}}, \bibinfo {author} {\bibfnamefont {J.}~\bibnamefont {Eppehimer}},
  \bibinfo {author} {\bibfnamefont {M.}~\bibnamefont {Skold}}, \bibinfo
  {author} {\bibfnamefont {M.}~\bibnamefont {Karaoulis}}, \bibinfo {author}
  {\bibfnamefont {L.}~\bibnamefont {Godinez}}, \ and\ \bibinfo {author}
  {\bibfnamefont {M.}~\bibnamefont {Prasad}},\ }\href@noop {} {\bibfield
  {journal} {\bibinfo  {journal} {Journal of Colloid and Interface Science}\
  }\textbf {\bibinfo {volume} {398}} (\bibinfo {year} {2013})}\BibitemShut
  {NoStop}%
\bibitem [{\citenamefont {Canan}(1999)}]{BC99}%
  \BibitemOpen
  \bibfield  {author} {\bibinfo {author} {\bibfnamefont {B.}~\bibnamefont
  {Canan}},\ }\emph {\bibinfo {title} {Dielectric Properties of Mixtures of
  Clay-Water-Organic Compounds}},\ \href@noop {} {Ph.D. thesis},\ \bibinfo
  {school} {Colorado School of Mines} (\bibinfo {year} {1999})\BibitemShut
  {NoStop}%
\bibitem [{\citenamefont {Weiler}\ and\ \citenamefont
  {Chaussidon}(1968)}]{WR68}%
  \BibitemOpen
  \bibfield  {author} {\bibinfo {author} {\bibfnamefont {R.~A.}\ \bibnamefont
  {Weiler}}\ and\ \bibinfo {author} {\bibfnamefont {J.}~\bibnamefont
  {Chaussidon}},\ }\href@noop {} {\bibfield  {journal} {\bibinfo  {journal}
  {Clay and Clay Minerals}\ }\textbf {\bibinfo {volume} {25}},\ \bibinfo
  {pages} {147 } (\bibinfo {year} {1968})}\BibitemShut {NoStop}%
\bibitem [{\citenamefont {Calvet}(1975)}]{RC75}%
  \BibitemOpen
  \bibfield  {author} {\bibinfo {author} {\bibfnamefont {R.}~\bibnamefont
  {Calvet}},\ }\href@noop {} {\bibfield  {journal} {\bibinfo  {journal} {Clay
  and Clay Minerals}\ }\textbf {\bibinfo {volume} {23}} (\bibinfo {year}
  {1975})}\BibitemShut {NoStop}%
\bibitem [{\citenamefont {Fripiat}\ \emph {et~al.}(1965)\citenamefont
  {Fripiat}, \citenamefont {Jelli}, \citenamefont {Poncelet},\ and\
  \citenamefont {Andre}}]{FJ65}%
  \BibitemOpen
  \bibfield  {author} {\bibinfo {author} {\bibfnamefont {J.~J.}\ \bibnamefont
  {Fripiat}}, \bibinfo {author} {\bibfnamefont {A.}~\bibnamefont {Jelli}},
  \bibinfo {author} {\bibfnamefont {G.}~\bibnamefont {Poncelet}}, \ and\
  \bibinfo {author} {\bibfnamefont {J.}~\bibnamefont {Andre}},\ }\href@noop {}
  {\bibfield  {journal} {\bibinfo  {journal} {J. Phys. chem.}\ }\textbf
  {\bibinfo {volume} {69}} (\bibinfo {year} {1965})}\BibitemShut {NoStop}%
\bibitem [{\citenamefont {Janek}\ \emph {et~al.}(2009)\citenamefont {Janek},
  \citenamefont {Bug\'{a}r}, \citenamefont {Lorenc}, \citenamefont {Sz\"{o}cs},
  \citenamefont {Veli\v{c}},\ and\ \citenamefont {Chorv\'{a}t}}]{MJ09}%
  \BibitemOpen
  \bibfield  {author} {\bibinfo {author} {\bibfnamefont {M.}~\bibnamefont
  {Janek}}, \bibinfo {author} {\bibfnamefont {I.}~\bibnamefont {Bug\'{a}r}},
  \bibinfo {author} {\bibfnamefont {D.}~\bibnamefont {Lorenc}}, \bibinfo
  {author} {\bibfnamefont {V.}~\bibnamefont {Sz\"{o}cs}}, \bibinfo {author}
  {\bibfnamefont {D.}~\bibnamefont {Veli\v{c}}}, \ and\ \bibinfo {author}
  {\bibfnamefont {D.}~\bibnamefont {Chorv\'{a}t}},\ }\href@noop {} {\bibfield
  {journal} {\bibinfo  {journal} {Clay and Clay Minerals}\ }\textbf {\bibinfo
  {volume} {57}},\ \bibinfo {pages} {416 } (\bibinfo {year}
  {2009})}\BibitemShut {NoStop}%
\bibitem [{\citenamefont {Scales}\ and\ \citenamefont
  {Batzle}(2006{\natexlab{a}})}]{JS0689}%
  \BibitemOpen
  \bibfield  {author} {\bibinfo {author} {\bibfnamefont {J.~A.}\ \bibnamefont
  {Scales}}\ and\ \bibinfo {author} {\bibfnamefont {M.}~\bibnamefont
  {Batzle}},\ }\href@noop {} {\bibfield  {journal} {\bibinfo  {journal}
  {Applied Physics Letters}\ }\textbf {\bibinfo {volume} {89}} (\bibinfo {year}
  {2006}{\natexlab{a}})}\BibitemShut {NoStop}%
\bibitem [{\citenamefont {Scales}\ and\ \citenamefont
  {Batzle}(2006{\natexlab{b}})}]{JS0688}%
  \BibitemOpen
  \bibfield  {author} {\bibinfo {author} {\bibfnamefont {J.~A.}\ \bibnamefont
  {Scales}}\ and\ \bibinfo {author} {\bibfnamefont {M.}~\bibnamefont
  {Batzle}},\ }\href@noop {} {\bibfield  {journal} {\bibinfo  {journal}
  {Applied Physics Letters}\ }\textbf {\bibinfo {volume} {88}} (\bibinfo {year}
  {2006}{\natexlab{b}})}\BibitemShut {NoStop}%
\bibitem [{\citenamefont {Greeney}\ and\ \citenamefont {Scales}(2012)}]{NG14}%
  \BibitemOpen
  \bibfield  {author} {\bibinfo {author} {\bibfnamefont {N.~S.}\ \bibnamefont
  {Greeney}}\ and\ \bibinfo {author} {\bibfnamefont {J.~A.}\ \bibnamefont
  {Scales}},\ }\href@noop {} {\bibfield  {journal} {\bibinfo  {journal}
  {Applied Physics Letters}\ }\textbf {\bibinfo {volume} {100}} (\bibinfo
  {year} {2012})}\BibitemShut {NoStop}%
\bibitem [{\citenamefont {Weiss}\ \emph {et~al.}(2009)\citenamefont {Weiss},
  \citenamefont {Zadler}, \citenamefont {Schafer},\ and\ \citenamefont
  {Scales}}]{MW09}%
  \BibitemOpen
  \bibfield  {author} {\bibinfo {author} {\bibfnamefont {M.}~\bibnamefont
  {Weiss}}, \bibinfo {author} {\bibfnamefont {B.}~\bibnamefont {Zadler}},
  \bibinfo {author} {\bibfnamefont {S.}~\bibnamefont {Schafer}}, \ and\
  \bibinfo {author} {\bibfnamefont {J.~A.}\ \bibnamefont {Scales}},\
  }\href@noop {} {\bibfield  {journal} {\bibinfo  {journal} {Journal of Applied
  Physics}\ }\textbf {\bibinfo {volume} {106}} (\bibinfo {year}
  {2009})}\BibitemShut {NoStop}%
\bibitem [{\citenamefont {Rahman}\ \emph {et~al.}(2013)\citenamefont {Rahman},
  \citenamefont {Taylor},\ and\ \citenamefont {Scales}}]{RR13a}%
  \BibitemOpen
  \bibfield  {author} {\bibinfo {author} {\bibfnamefont {R.}~\bibnamefont
  {Rahman}}, \bibinfo {author} {\bibfnamefont {P.~C.}\ \bibnamefont {Taylor}},
  \ and\ \bibinfo {author} {\bibfnamefont {J.~A.}\ \bibnamefont {Scales}},\
  }\href@noop {} {\bibfield  {journal} {\bibinfo  {journal} {Review of
  Scientific Instruments}\ }\textbf {\bibinfo {volume} {84}} (\bibinfo {year}
  {2013})}\BibitemShut {NoStop}%
\bibitem [{\citenamefont {Dudorov}\ \emph {et~al.}(2005)\citenamefont
  {Dudorov}, \citenamefont {Lioubtchenko}, \citenamefont {Mallat},\ and\
  \citenamefont {Raisanen}}]{Dudorov95}%
  \BibitemOpen
  \bibfield  {author} {\bibinfo {author} {\bibfnamefont {S.~N.}\ \bibnamefont
  {Dudorov}}, \bibinfo {author} {\bibfnamefont {D.~V.}\ \bibnamefont
  {Lioubtchenko}}, \bibinfo {author} {\bibfnamefont {J.~A.}\ \bibnamefont
  {Mallat}}, \ and\ \bibinfo {author} {\bibfnamefont {A.~V.}\ \bibnamefont
  {Raisanen}},\ }\href@noop {} {\bibfield  {journal} {\bibinfo  {journal}
  {{IEEE} Transaction On Instrumentation and Measurement}\ }\textbf {\bibinfo
  {volume} {54}},\ \bibinfo {pages} {1916} (\bibinfo {year}
  {2005})}\BibitemShut {NoStop}%
\bibitem [{\citenamefont {Cullen}\ and\ \citenamefont {Yu}(1971)}]{cullen1}%
  \BibitemOpen
  \bibfield  {author} {\bibinfo {author} {\bibfnamefont {A.~L.}\ \bibnamefont
  {Cullen}}\ and\ \bibinfo {author} {\bibfnamefont {P.~K.}\ \bibnamefont
  {Yu}},\ }\href@noop {} {\bibfield  {journal} {\bibinfo  {journal}
  {Proceedings of The Royal Society A}\ }\textbf {\bibinfo {volume} {325}},\
  \bibinfo {pages} {49} (\bibinfo {year} {1971})}\BibitemShut {NoStop}%
\bibitem [{\citenamefont {Yu}\ and\ \citenamefont {Cullen}(1982)}]{cullen2}%
  \BibitemOpen
  \bibfield  {author} {\bibinfo {author} {\bibfnamefont {P.~K.}\ \bibnamefont
  {Yu}}\ and\ \bibinfo {author} {\bibfnamefont {A.~L.}\ \bibnamefont
  {Cullen}},\ }\href@noop {} {\bibfield  {journal} {\bibinfo  {journal}
  {Proceedings of The Royal Society A}\ }\textbf {\bibinfo {volume} {380}},\
  \bibinfo {pages} {49} (\bibinfo {year} {1982})}\BibitemShut {NoStop}%
\bibitem [{\citenamefont {Hirovnen}\ \emph {et~al.}(1996)\citenamefont
  {Hirovnen}, \citenamefont {Vainikainen}, \citenamefont {Lozowski},\ and\
  \citenamefont {Raisanen}}]{TH96}%
  \BibitemOpen
  \bibfield  {author} {\bibinfo {author} {\bibfnamefont {M.~T.}\ \bibnamefont
  {Hirovnen}}, \bibinfo {author} {\bibfnamefont {P.}~\bibnamefont
  {Vainikainen}}, \bibinfo {author} {\bibfnamefont {A.}~\bibnamefont
  {Lozowski}}, \ and\ \bibinfo {author} {\bibfnamefont {A.~V.}\ \bibnamefont
  {Raisanen}},\ }\href@noop {} {\bibfield  {journal} {\bibinfo  {journal} {IEEE
  Transaction on Instrumentation and Measurement}\ }\textbf {\bibinfo {volume}
  {45}},\ \bibinfo {pages} {780} (\bibinfo {year} {1996})}\BibitemShut
  {NoStop}%
\bibitem [{\citenamefont {Strom}\ and\ \citenamefont {Taylor}(1977)}]{ST77}%
  \BibitemOpen
  \bibfield  {author} {\bibinfo {author} {\bibfnamefont {U.}~\bibnamefont
  {Strom}}\ and\ \bibinfo {author} {\bibfnamefont {P.~C.}\ \bibnamefont
  {Taylor}},\ }\href@noop {} {\bibfield  {journal} {\bibinfo  {journal} {Phys.
  Rev. B}\ }\textbf {\bibinfo {volume} {16}},\ \bibinfo {pages} {5512}
  (\bibinfo {year} {1977})}\BibitemShut {NoStop}%
\bibitem [{\citenamefont {Jackson}(1985)}]{JML85}%
  \BibitemOpen
  \bibfield  {author} {\bibinfo {author} {\bibfnamefont {M.}~\bibnamefont
  {Jackson}},\ }\href@noop {} {\emph {\bibinfo {title} {Soil Chemical Analysis,
  Advanced Course}}},\ \bibinfo {edition} {2nd}\ ed.\ (\bibinfo  {publisher}
  {Published by the author, Madison, WI},\ \bibinfo {year} {1985})\BibitemShut
  {NoStop}%
\bibitem [{\citenamefont {Moore}\ and\ \citenamefont {Reynolds}(1997)}]{MDM97}%
  \BibitemOpen
  \bibfield  {author} {\bibinfo {author} {\bibfnamefont {D.}~\bibnamefont
  {Moore}}\ and\ \bibinfo {author} {\bibfnamefont {R.~J.}\ \bibnamefont
  {Reynolds}},\ }\href@noop {} {\emph {\bibinfo {title} {X-ray Diffraction and
  the Identification and Analysis of Clay Minerals}}},\ \bibinfo {edition}
  {2nd}\ ed.\ (\bibinfo  {publisher} {Oxford University Press, New York},\
  \bibinfo {year} {1997})\BibitemShut {NoStop}%
\bibitem [{\citenamefont {J}\ \emph {et~al.}(2007)\citenamefont {J},
  \citenamefont {P},\ and\ \citenamefont {F.}}]{Rouquerol2007}%
  \BibitemOpen
  \bibfield  {author} {\bibinfo {author} {\bibfnamefont {R.}~\bibnamefont {J}},
  \bibinfo {author} {\bibfnamefont {L.}~\bibnamefont {P}}, \ and\ \bibinfo
  {author} {\bibfnamefont {R.}~\bibnamefont {F.}},\ }\href@noop {} {\bibfield
  {journal} {\bibinfo  {journal} {Studies in Surface Science and Catalysis}\ ,\
  \bibinfo {pages} {49 }} (\bibinfo {year} {2007})}\BibitemShut {NoStop}%
\bibitem [{\citenamefont {Kuila}\ \emph {et~al.}(2014)\citenamefont {Kuila},
  \citenamefont {McCarty}, \citenamefont {Derkowski}, \citenamefont {Fischer},
  \citenamefont {Top\'{o}r},\ and\ \citenamefont {Prasad}}]{Kuila2014}%
  \BibitemOpen
  \bibfield  {author} {\bibinfo {author} {\bibfnamefont {U.}~\bibnamefont
  {Kuila}}, \bibinfo {author} {\bibfnamefont {D.}~\bibnamefont {McCarty}},
  \bibinfo {author} {\bibfnamefont {A.}~\bibnamefont {Derkowski}}, \bibinfo
  {author} {\bibfnamefont {T.}~\bibnamefont {Fischer}}, \bibinfo {author}
  {\bibfnamefont {T.}~\bibnamefont {Top\'{o}r}}, \ and\ \bibinfo {author}
  {\bibfnamefont {M.}~\bibnamefont {Prasad}},\ }\href@noop {} {\bibfield
  {journal} {\bibinfo  {journal} {Fuel}\ }\textbf {\bibinfo {volume} {135}},\
  \bibinfo {pages} {359} (\bibinfo {year} {2014})}\BibitemShut {NoStop}%
\bibitem [{\citenamefont {Drits}\ and\ \citenamefont
  {Sakharov}(1976)}]{drits1976}%
  \BibitemOpen
  \bibfield  {author} {\bibinfo {author} {\bibfnamefont {V.}~\bibnamefont
  {Drits}}\ and\ \bibinfo {author} {\bibfnamefont {B.}~\bibnamefont
  {Sakharov}},\ }\href@noop {} {\emph {\bibinfo {title} {X-ray Analysis of
  Mixed-layer Clay Minerals}}}\ (\bibinfo  {publisher} {Nauka},\ \bibinfo
  {address} {Moscow},\ \bibinfo {year} {1976})\BibitemShut {NoStop}%
\bibitem [{\citenamefont {Drits}\ \emph {et~al.}(1997)\citenamefont {Drits},
  \citenamefont {Sakharov}, \citenamefont {Lindgreen},\ and\ \citenamefont
  {Salyn}}]{drits1997}%
  \BibitemOpen
  \bibfield  {author} {\bibinfo {author} {\bibfnamefont {V.}~\bibnamefont
  {Drits}}, \bibinfo {author} {\bibfnamefont {B.}~\bibnamefont {Sakharov}},
  \bibinfo {author} {\bibfnamefont {H.}~\bibnamefont {Lindgreen}}, \ and\
  \bibinfo {author} {\bibfnamefont {A.}~\bibnamefont {Salyn}},\ }\href@noop {}
  {\bibfield  {journal} {\bibinfo  {journal} {Clay Minerals}\ }\textbf
  {\bibinfo {volume} {32}},\ \bibinfo {pages} {351} (\bibinfo {year}
  {1997})}\BibitemShut {NoStop}%
\bibitem [{\citenamefont {Sakharov}\ \emph {et~al.}(1999)\citenamefont
  {Sakharov}, \citenamefont {Lindgreen}, \citenamefont {Salyn},\ and\
  \citenamefont {Drits}}]{Sahkarov1999}%
  \BibitemOpen
  \bibfield  {author} {\bibinfo {author} {\bibfnamefont {B.}~\bibnamefont
  {Sakharov}}, \bibinfo {author} {\bibfnamefont {H.}~\bibnamefont {Lindgreen}},
  \bibinfo {author} {\bibfnamefont {A.}~\bibnamefont {Salyn}}, \ and\ \bibinfo
  {author} {\bibfnamefont {V.}~\bibnamefont {Drits}},\ }\href@noop {}
  {\bibfield  {journal} {\bibinfo  {journal} {Clays and Clay Minerals}\
  }\textbf {\bibinfo {volume} {47}},\ \bibinfo {pages} {555} (\bibinfo {year}
  {1999})}\BibitemShut {NoStop}%
\bibitem [{\citenamefont {McCarty}\ \emph {et~al.}(2009)\citenamefont
  {McCarty}, \citenamefont {Sakharov},\ and\ \citenamefont
  {Drits}}]{McCarty2009}%
  \BibitemOpen
  \bibfield  {author} {\bibinfo {author} {\bibfnamefont {D.}~\bibnamefont
  {McCarty}}, \bibinfo {author} {\bibfnamefont {B.}~\bibnamefont {Sakharov}}, \
  and\ \bibinfo {author} {\bibfnamefont {V.}~\bibnamefont {Drits}},\
  }\href@noop {} {\bibfield  {journal} {\bibinfo  {journal} {American
  Mineralogist}\ }\textbf {\bibinfo {volume} {94}},\ \bibinfo {pages} {1653}
  (\bibinfo {year} {2009})}\BibitemShut {NoStop}%
\bibitem [{\citenamefont {Zallen}(1983)}]{ZLN83}%
  \BibitemOpen
  \bibfield  {author} {\bibinfo {author} {\bibfnamefont {R.}~\bibnamefont
  {Zallen}},\ }\href@noop {} {\emph {\bibinfo {title} {The Physics of Amorphous
  Solids}}}\ (\bibinfo  {publisher} {Wiley-Inerscience},\ \bibinfo {year}
  {1983})\BibitemShut {NoStop}%
\bibitem [{\citenamefont {Rahman}\ \emph {et~al.}(2014)\citenamefont {Rahman},
  \citenamefont {Ohno}, \citenamefont {Taylor},\ and\ \citenamefont
  {Scales}}]{RR14}%
  \BibitemOpen
  \bibfield  {author} {\bibinfo {author} {\bibfnamefont {R.}~\bibnamefont
  {Rahman}}, \bibinfo {author} {\bibfnamefont {T.~R.}\ \bibnamefont {Ohno}},
  \bibinfo {author} {\bibfnamefont {P.~C.}\ \bibnamefont {Taylor}}, \ and\
  \bibinfo {author} {\bibfnamefont {J.~A.}\ \bibnamefont {Scales}},\
  }\href@noop {} {\bibfield  {journal} {\bibinfo  {journal} {Appl. Phys.
  Lett.}\ }\textbf {\bibinfo {volume} {104}} (\bibinfo {year}
  {2014})}\BibitemShut {NoStop}%
\bibitem [{\citenamefont {Debye}(1929)}]{PD29}%
  \BibitemOpen
  \bibfield  {author} {\bibinfo {author} {\bibfnamefont {P.}~\bibnamefont
  {Debye}},\ }\href@noop {} {\emph {\bibinfo {title} {Polar Molecules}}}\
  (\bibinfo  {publisher} {Chemical Catalogue Company, NY},\ \bibinfo {year}
  {1929})\BibitemShut {NoStop}%
\bibitem [{\citenamefont {Hamon}(1953)}]{BH53}%
  \BibitemOpen
  \bibfield  {author} {\bibinfo {author} {\bibfnamefont {B.~V.}\ \bibnamefont
  {Hamon}},\ }\href@noop {} {\bibfield  {journal} {\bibinfo  {journal}
  {Australian Journal of Physics}\ }\textbf {\bibinfo {volume} {6}},\ \bibinfo
  {pages} {304} (\bibinfo {year} {1953})}\BibitemShut {NoStop}%
\bibitem [{\citenamefont {Seleznev}\ \emph {et~al.}(2011)\citenamefont
  {Seleznev} \emph {et~al.}}]{NS11}%
  \BibitemOpen
  \bibfield  {author} {\bibinfo {author} {\bibfnamefont {N.~V.}\ \bibnamefont
  {Seleznev}} \emph {et~al.},\ }in\ \href@noop {} {\emph {\bibinfo {booktitle}
  {SPWLA 52nd Annual Logging Symposium, Colorado Springs, CO, USA, May,
  2011}}}\ (\bibinfo  {publisher} {SPWLA},\ \bibinfo {year} {2011})\BibitemShut
  {NoStop}%
\bibitem [{\citenamefont {Egashira}\ and\ \citenamefont
  {Matsumoto}(1981)}]{KE81}%
  \BibitemOpen
  \bibfield  {author} {\bibinfo {author} {\bibfnamefont {K.}~\bibnamefont
  {Egashira}}\ and\ \bibinfo {author} {\bibfnamefont {J.}~\bibnamefont
  {Matsumoto}},\ }\href@noop {} {\bibfield  {journal} {\bibinfo  {journal}
  {Soil. Sci. Plant Nutr.}\ }\textbf {\bibinfo {volume} {27}},\ \bibinfo
  {pages} {289 } (\bibinfo {year} {1981})}\BibitemShut {NoStop}%
\bibitem [{\citenamefont {Zadaka}\ \emph {et~al.}(2010)\citenamefont {Zadaka},
  \citenamefont {Radian},\ and\ \citenamefont {Mishael}}]{DZ10}%
  \BibitemOpen
  \bibfield  {author} {\bibinfo {author} {\bibfnamefont {D.}~\bibnamefont
  {Zadaka}}, \bibinfo {author} {\bibfnamefont {A.}~\bibnamefont {Radian}}, \
  and\ \bibinfo {author} {\bibfnamefont {Y.}~\bibnamefont {Mishael}},\
  }\href@noop {} {\bibfield  {journal} {\bibinfo  {journal} {Journal of Colloid
  and Interface Science}\ }\textbf {\bibinfo {volume} {352}},\ \bibinfo {pages}
  {171 } (\bibinfo {year} {2010})}\BibitemShut {NoStop}%
\end{thebibliography}%

\end{document}